\documentclass[twocolumn,showpacs,aps,prb,preprintnumbers,amsmath,amssymb]{revtex4}

\usepackage{graphicx}
\usepackage{dcolumn}
\usepackage{bm}
\usepackage{epsfig}

\begin{document}


\title{
Dirac fermion quantization on graphene edges: \\
Isospin-orbit coupling, zero modes and spontaneous valley polarization
}

\author{Grigory Tkachov}
\affiliation{
Max Planck Institute for the Physics of Complex Systems, Dresden, Germany
}

\begin{abstract}
The paper addresses boundary electronic properties of graphene 
with a complex edge structure of the armchair/zigzag/armchair type. 
It is shown that the finite zigzag region supports edge bound states 
with discrete equidistant spectrum obtained from 
the Green's function of the continuum Dirac equation. 
The energy levels exhibit the coupling between 
the valley degree of freedom and the orbital quantum number, 
analogous to a spin-orbit interaction.
The characteristic feature of the spectrum is 
the presence of a zero mode, the bound state of vanishing energy. 
It resides only in one of the graphene valleys,  
breaking spontaneously Kramers' symmetry of the edge states. 
This implies the spontaneous valley polarization 
characterized by the valley isospin $\pm 1/2$.   
The polarization is manifested by a zero-magnetic field anomaly 
in the local tunneling density of states, and 
is directly related to the local electric Hall conductivity.
\end{abstract}

\pacs{73.20.At,73.22.Gk,73.63.Bd}

\maketitle

\section{Introduction}
\label{intro}

Due to the close connection between their topological and physical properties,
two-dimensional (2D) electron systems have traditionally been in the focus of fundamental research. 
From the practical side, device functionalities in the 2D geometry 
are of great importance for applications and 
particularly suitable for lateral electronic architecture.   
The interest in these general aspects of 2D electron systems 
has recently revived in the light of the experimental success 
in isolating individual layers of graphite, preserving the honeycomb crystal structure~\cite{Novoselov05,Zhang05}. 
Such a system - graphene - exhibits elementary excitations 
behaving at low energies and long distances 
as massless Dirac fermions~\cite{Wallace47,Semenoff84}.  
Due to its massless quasiparticles  
graphene stands out among other 2D electron systems,  
which is probably most prominently manifested by 
the unconventional quantum Hall physics (e.g. Refs.~\onlinecite{Novoselov05,Zhang05,Gusynin05,McCann06,Nomura06,Abanin07}), 
the phenomenon of Klein tunneling~\cite{Klein} 
and fermion bound states on extended defects such as graphene boundaries~\cite{Fujita96,Waka00,Koba05,Niimi06,Zhou06,Peres06,Brey06,GT07,Akhmerov08,Castro08,GT08}, to name a few.
In particular,  understanding boundary effects in clean and disordered~\cite{Evaldsson08,Mucciolo08} 
graphene and the need for their characterization are 
among the outstanding current challenges in the field, 
arising from potentially promising electronic applications 
of graphene ribbons~\cite{Han07,Chen07} and quantum dots~\cite{Ponomarenko08}.

\begin{figure}[t]
\begin{center}
\epsfxsize=1\hsize
\epsffile{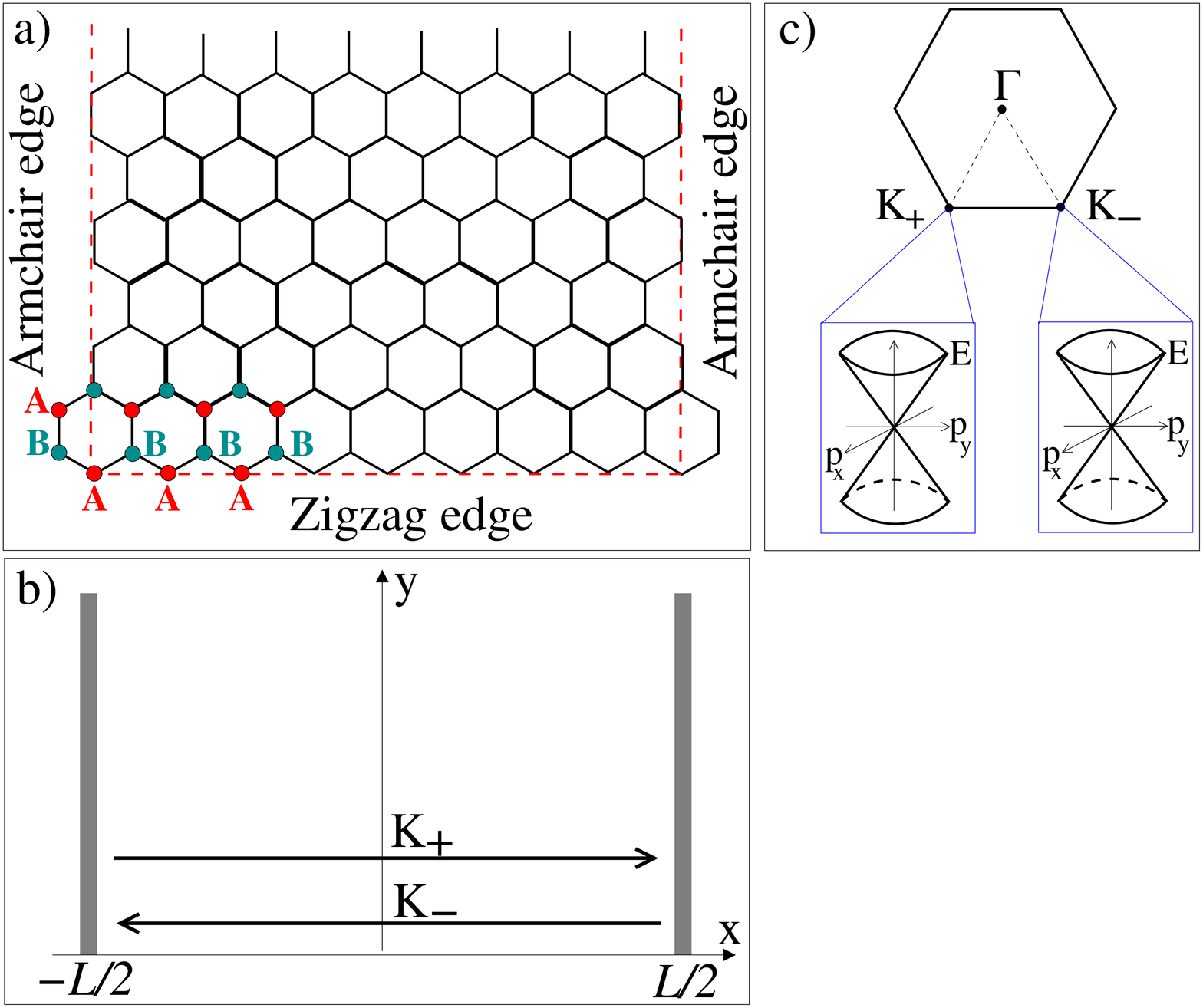}
\end{center}
\caption{ (Color online)
{\bf (a)} Example of a finite-length zigzag edge sided by two armchair boundaries. 
$A$ and $B$ mark the sites of the two triangular sublattices.
{\bf (b)} Geometry of the continuum model for the system in panel (a): 
The zigzag edge at $y=0$ supports a Kramers' pair of counter-propagating edge states from $K_+$ and $K_-$ valleys.
They transform into each other via intervalley scattering caused by the armchair sides at $x=\pm L/2$. 
{\bf (c)} Schematic view of the spectrum near the nodal points $K_+$ and $K_-$ of graphene's Brillouin zone~\cite{Wallace47}.
}
\label{Geo}
\end{figure}

One of the reasons why the boundary effects in graphene should matter
was pointed out quite a time ago by Fujita et al [Ref.~\onlinecite{Fujita96}]. 
Using tight-binding calculations they predicted 
a new branch of quasiparticle states 
localized on the so-called "zigzag" edge. 
It is one of the most common types of the honeycomb lattice termination 
formed by two parallel crystal faces of the triangular sublattices 
of the honeycomb structure [see,  Fig.~\ref{Geo}(a)].
The properties of the zigzag edge states are better understood 
when compared to the edge states in conventional 2D quantum Hall systems~\cite{Halperin82}. 
Unlike the latter, the zigzag edge states exist without any external magnetic field  and  any excitation gap in the 2D bulk.
They are nonchiral: there is a Kramers' pair of counter-propagating modes 
originating from two nonequivalent nodal points of graphene's Brillouin zone [see, Figs.~\ref{Geo}(b) and (c)]. 
The zigzag edge states have essentially the same origin as the bound states of massless fermions on domain walls~\cite{Jackiw76}. 
Here the role of the domain wall is assumed by  the out-of-plane rotation 
of the "sublattice" spin which in the continuum limit 
corresponds to the zigzag edge~\cite{GT08}. 
Experimental evidence for the bound states on graphene edges comes from 
both tunneling~\cite{Koba05,Niimi06} and 
angle-resolved photoemission spectroscopies~\cite{Zhou06}.

The present study is motivated by the observation that in experiments one has to deal with finite-length zigzag edges 
that represent a section of the graphene boundary sided usually 
by two armchair edges~\cite{Koba05,Niimi06}.
As the armchair sides do not support edge states~\cite{Fujita96,Brey06}, 
one should generally expect quantization of the propagating modes 
in the finite zigzag section. 
This type of quantization is distinct from the size-quantization in zigzag graphene ribbons studied earlier~\cite{Fujita96,Waka00,Peres06,Brey06},
because it can occur on an {\em isolated}  zigzag boundary, 
which is the typical situation in scanning tunneling experiments~\cite{Koba05,Niimi06}.  
The consequences of such a quantization have not been studied previously. 
In the present work, they are addressed 
within the Dirac fermion confinement model 
derived from the lattice structure shown in Fig.~\ref{Geo}(a).

In our approach the time-reversal symmetry and  Kramers' degeneracy of the zigzag edge states 
comes as a result of an effective isospin-orbit coupling. 
The isospin is introduced as a convenient formal representation for the two nonequivalent nodal points of graphene's Brillouin zone. 
The rotations generated by the isospin leave the 2D Dirac equation invariant.
We show that this continuous symmetry is broken by the zigzag confinement, and  
the edge state spectrum explicitly depends on the confinement parameters controlling the isospin-orbit coupling. 
The quantization of the edge states is achieved by imposing effective boundary conditions at the ends of the zigzag edge [see, Fig.~\ref{Geo}(b)].
They cause the intervalley scattering connecting the incident and outgoing edge states, which models the armchair confinement. 
It turns out that the quantized spectrum contains a zero mode, i.e. the state with vanishing momentum and energy. 
Remarkably, it couples only to one of the isospin projections, that is it exists only in one of the valleys, 
breaking spontaneously the  Kramers' symmetry of the edge states.  
This leads to the spontaneous isospin (valley) polarization with the total edge-state isospin $\pm 1/2$. 
This mechanism of the valley polarization differs from the previous proposals~\cite{Rycerz07}.
We demonstrate that the spontaneous symmetry breaking can be detected through the magnetic-field dependence of the tunneling density of states, 
and also find a direct relation between the isospin polarization and the local electric Hall conductivity.

The subsequent sections give a complete account of our approach: 
In Sec.~\ref{bp} we formulate the boundary problem for a finite zigzag edge and 
analyze it in terms of the discrete and continuous symmetries of the 2D Dirac fermions. 
The Green's function solution of the boundary problem and 
the spectrum of the quantized Dirac fermion edge states are discussed in Sec.~\ref{finite}.  
Section~\ref{valley} addresses the valley polarization effects,  
both spontaneous and induced. 
The latter is the analogue of the quantum spin Hall polarization.  
Finally, section~\ref{signs} describes the signatures of the valley polarization in observables, 
such as the tunneling density of states and the local electric Hall conductivity, and contains concluding discussion. 
 
\section{The boundary problem}
\label{bp}

\subsection{2D massless fermions, chiral symmetry and isospin}
\label{general}

The two distinct nodal points (valleys) of graphene's Brillouin zone 
result in a pair of massless Weyl fermions  
whose wave functions, $\psi_+$ and $\psi_-$, 
satisfy the matrix equation:  
\begin{eqnarray}
&
\epsilon\Psi=H\Psi, 
\Psi=\left[
\begin{array}{c}
\psi_+ \\
\psi_- 
\end{array}
\right],
H=v\left[
\begin{array}{cc}
\mbox{\boldmath$\sigma$}{\bf p}  &  0\\
0 &  U(\mbox{\boldmath$\sigma$}{\bf p})U^{-1}
\end{array}
\right].
&
\label{D}
\end{eqnarray}
It is assumed that the Hamiltonian $H$ is 
diagonal in valley space $(+,-)$. 
The intra-valley Hamiltonians are expressed in terms 
of the Pauli matrices $\sigma_{x,y,z}$ acting on the functions 
\begin{equation}
  \psi_\pm =
\left[
\begin{array}{c}
\psi_{A_\pm} \\
\psi_{B_\pm}
\end{array}
\right],
\label{PsiAB}
\end{equation}
that have two components due to the bipartite lattice structure of graphene, 
with two sublattices denoted as  $A$ and $B$ in Fig.~\ref{Geo}(a);
$v$ and $\epsilon$ are the Fermi velocity and energy 
with respect to the Fermi level, and 
the quasiparticle momentum ${\bf p}$ 
is confined to the plane of the system.

We further assume that the intra-valley Hamiltonians 
are related to each other by the chiral symmetry:
\begin{eqnarray}
U\,(\mbox{\boldmath$\sigma$}{\bf p})\,U^{-1}= -\mbox{\boldmath$\sigma$}{\bf p},
\label{chiral}
\end{eqnarray}
where $U$ is a unitary matrix.
In this way we explicitly account for the generic property of 
nodal lattice quasiparticles known as fermion doubling: 
they come in pairs of opposite-chirality (Weyl) species 
that together obey the Dirac equation~\cite{NN81}.   
We note that in the 2D case the unitary transformation, Eq.~(\ref{chiral}) 
is always achieved by one of the $\sigma$ matrices. 
If, for instance, the system is located in the $x,y$ plane [Fig.~\ref{Geo}],   
we have
\begin{eqnarray}
\mbox{\boldmath$\sigma$}{\bf p}=\sigma_x p_x + \sigma_y p_y,
\qquad 
U=\sigma_z.
\label{U}
\end{eqnarray}

The discrete chiral symmetry, Eq.~(\ref{chiral}) 
can be promoted to a continuous one. 
Let us introduce another set of the Pauli matrices $\tau_{1,2,3}$,
acting in the valley space, and consider the vector operator,
\begin{eqnarray}
	{\bf I}=\frac{1}{2}\,
	\biggl(\tau_1\otimes\sigma_z, \tau_2\otimes\sigma_z, \tau_3\otimes\sigma_0
	\biggr), 
	\,\,
	[I_k,I_l]=i\varepsilon_{klm}I_m,
	\label{I}
\end{eqnarray}
whose components $I_k$ ($k=1,2,3$) formally satisfy 
the commutation relations of an angular momentum. 
It is easy to see that the Hamiltonian $H$ is invariant 
under rotations generated by $I_k$:   
\begin{eqnarray}
	{\rm e}^{i\theta_k I_k}H{\rm e}^{-i\theta_k I_k}=H,\quad 
	H=v\tau_3\otimes(\mbox{\boldmath$\sigma$}{\bf p}), 
	\quad 
	\label{Rot}
\end{eqnarray}
where $\theta_k$ is the rotation angle.
This means that the original choice of the upper and lower components 
of $\Psi$ [Eq.~(\ref{D})]
as being the "$+$" and "$-$" valley functions, respectively, 
is not physically distinguished. 
One can rather treat them as the "up" and "down" states 
of the effective spin (isospin) $1/2$. 
We will nevertheless keep the original notations $\psi_\pm$ 
for the upper and lower components of $\Psi$, 
interpreting them as the projections
\begin{eqnarray}
	\psi_\pm=\left(\frac{I_0}{2} \pm I_3\right)\Psi,
	\label{project} 
\end{eqnarray}
where $I_0=\tau_0\otimes\sigma_0$ is the $4\times 4$ unit matrix 
(the direct product of the $2\times 2$ unit matrices $\tau_0$ and $\sigma_0$).

\subsection{
Boundary condition for the zigzag edge and 
broken isospin rotation symmetry}
\label{zig}

The zigzag edge is a type of the honeycomb lattice termination
where the outermost lattice sites all belong to one of the sublattices [Fig.~\ref{Geo}(a)]. 
It does not couple the states from the $K_+$ and $K_-$ valleys~\cite{Fujita96}, 
due to which the continuum boundary condition for the zigzag edge can be obtained by rather simple reasoning.~\cite{Brey06} 
To be concrete let us assume that the outermost sites are all of the $A$ type and the next (missing) 
atomic row would be of the $B$ type, as in Fig.~\ref{Geo}(a). 
On the missing $B$ row one can impose the hard-wall condition $\psi_{B\pm}(x,0)=0$, 
while keeping  $\psi_{A\pm}(x,0)$ arbitrary. In spinor notations [Eq.~(\ref{PsiAB})], this reads 
\begin{equation}
  \psi_\pm(x,0)=\mbox{\boldmath$\sigma$}{\bf l}_\perp  \psi_\pm(x,0), \quad {\bf l}_\perp=(0,0,1).
\label{BCzig}
\end{equation} 
This boundary condition admits the generalization 
beyond the hard-wall approximation. 
It is achieved  by rotating the unit vector ${\bf l}_\perp$ 
about the normal ${\bf n}_B$ to the boundary 
(in Fig.~\ref{Geo}(b), ${\bf n}_B\|\, {\bf{\hat y}}$), 
which is consistent with the requirement for 
the normal component of the current to vanish at the edge~\cite{McCann04,Akhmerov08}. 
Moreover, the rotation can be made valley-dependent: 
${\bf l}_\perp\to {\bf l}_\pm$.
Using the $4$ component spinors, we can therefore write: 
\begin{equation}
	\Psi(x,0)=M\Psi(x,0),
\label{BC1}
\end{equation}
\begin{equation}
M=\frac{\tau_0 + \tau_3}{2}\otimes\mbox{\boldmath$\sigma$}{\bf l}_+  
	        + 
         \frac{\tau_0 - \tau_3}{2}\otimes\mbox{\boldmath$\sigma$}{\bf l}_- ,
\label{M}
\end{equation}
\begin{equation}
    {\bf l}^2_\pm =1, \qquad ({\bf l}_\pm {\bf n}_B)=0.
\label{n+-}
\end{equation}
Further restrictions on ${\bf l}_\pm$ are imposed by 
the discrete symmetries of the problem.
As the lattice prototype of our system has two identical sides [Fig.~\ref{Geo}(a)], 
our continuum model should inherit spatial parity with respect to coordinate reflection along the edge, i.e. $x\to -x$.  
It is the symmetry of the Dirac equation (\ref{D}) 
since the coordinate reflection can be compensated  by the spinor transformation:
\begin{eqnarray}
	\Psi^P(x,y)=\Lambda \Psi(-x,y), \quad \Lambda=\tau_1\otimes\sigma_x,
\label{parity}
\end{eqnarray}
simultaneously swapping both the valley and sublattice spinor components. 
However, the boundary condition, Eq.~(\ref{BC1}) does not share this symmetry 
because $M$ and $\Lambda$ do not commute
\begin{eqnarray}
	\Lambda M \Lambda^{-1} &=& 
	      \frac{\tau_0 - \tau_3}{2}\otimes ( \sigma_x l_{x+} - \sigma_z l_{z+} ) + 
       \nonumber\\
       & +& 
        \frac{\tau_0 + \tau_3}{2}\otimes ( \sigma_x l_{x-} - \sigma_z l_{z-} ), 
\label{Mprime}
\end{eqnarray}
unless there is a relation between ${\bf l}_+$ and  ${\bf l}_-$ such that 
\begin{eqnarray}
  l_{x+}=l_{x-}\equiv l_x, \quad  l_{z+}=-l_{z-}\equiv l_z, \quad 
  {\bf l}=(l_x, 0\,, l_z). 
 \label{n}
\end{eqnarray} 
These restrictions also make the zigzag boundary 
invariant under time-reversal operation 
$\Psi^{T}(x,y)=\Lambda \Psi^*(x,y)$.

We are now prepared to prove that the zigzag boundary condition, 
Eq.~(\ref{BC1}) violates the isospin rotation symmetry. 
More specifically, we are talking about the nontrivial rotations 
generated by the $I_1$ and $I_2$ components of the isospin, Eq.~(\ref{I}). 
Indeed, the matrix $M$ (\ref{M}) does not commute with $I_{1,2}$:
\begin{eqnarray}
	I_{1,2}MI^{-1}_{1,2}&=&
	       \frac{\tau_0 - \tau_3}{2}\otimes ( -\sigma_x l_{x+} + \sigma_z l_{z+} ) + 
       \nonumber\\
       & +& 
        \frac{\tau_0 + \tau_3}{2}\otimes ( - \sigma_x l_{x-} + \sigma_z l_{z-} ), 
\label{M_I1I2}
\end{eqnarray}
unless ${\bf l}_+$ and  ${\bf l}_-$ satisfy the conditions:
\begin{eqnarray}
  l_{x+}=-l_{x-}, \quad  l_{z+}=l_{z-}. 
 \label{n1}
\end{eqnarray} 
These are incompatible with the requirements for the $x$-parity and time-reversal symmetry, Eq.~(\ref{n}).
{\em In section~\ref{edge} we demonstrate 
that the broken isospin rotation symmetry 
implies an analogue of the spin-orbit coupling 
controlled by the components of the vector ${\bf l}$ in Eq.~(\ref{n}).}

\subsection{Parity-symmetric armchair edges}
\label{arm}

We now turn to the boundary conditions at the armchair sides $x=\pm L/2$. 
They should account for the valley and sublattice mixing 
specific to the armchair lattice termination~\cite{Fujita96} 
and, at the same time, possess both the $x$-parity and time-reversal symmetry.  
The suitable boundary conditions can be written as~\cite{Armchair}
\begin{eqnarray}
&
	\Psi\left(\pm \frac{L}{2},y\right)= \Lambda\Psi\left(\pm \frac{L}{2},y\right),
&
\label{BCarm}
\end{eqnarray}
with the same off-diagonal matrix $\Lambda$ as in Eq.~(\ref{parity}). 
They meet the requirement of the vanishing of the normal component 
of the Dirac current:  
\begin{eqnarray}
&
  j_x\left(\pm\frac{L}{2},y\right)=\Psi^\dagger\left(\pm\frac{L}{2},y\right)\tau_3\otimes\sigma_x\Psi\left(\pm\frac{L}{2},y\right)
&
\label{jx}\\
 &
=\Psi^\dagger\left(\pm\frac{L}{2},y\right)\tau_1\otimes\sigma_x (\tau_3\otimes\sigma_x) \tau_1\otimes\sigma_x\Psi\left(\pm\frac{L}{2},y\right)
&
\nonumber\\
& 
=-\Psi^\dagger\left(\pm\frac{L}{2},y\right)\tau_3\otimes\sigma_x\Psi\left(\pm\frac{L}{2},y\right)=-  j_x\left(\pm\frac{L}{2},y\right)=0,
&
\nonumber	
\end{eqnarray}
where we have switched to the creation $\Psi^\dagger(x,y)$ and 
annihilation $\Psi(x,y)$ operators. 

Importantly, the $x$-parity of the problem allows 
us to reduce the boundary conditions, 
Eq.~(\ref{BCarm}) to the usual symmetric boundary conditions:
\begin{eqnarray}
&
\Psi\left(\frac{L}{2},y\right)=\Psi\left(-\frac{L}{2},y\right).
&
\label{BCsym}
\end{eqnarray}
To prove this we first notice that the original function $\Psi(x,y)$ 
and the transformed one $\Psi^P(x,y)$ [Eq.~(\ref{parity})]
correspond to the same solution of Eqs.~(\ref{D}), (\ref{BC1}) and (\ref{BCarm}) ,
and, therefore, must coincide: $\Psi(x,y)=\Lambda \Psi(-x,y)$. 
In particular, at $x=\pm L/2$ we have 
\begin{eqnarray}
&
	\Psi\left(\pm\frac{L}{2},y\right)=\Lambda\Psi\left(\mp\frac{L}{2},y\right).
&
\label{parity1}
\end{eqnarray}
Comparison with Eq.~(\ref{BCarm}) yields Eq.~(\ref{BCsym}). 
For the actual calculations, we will use the symmetric boundary conditions 
modulated by a magnetic phase $\phi$:
\begin{eqnarray}
	&
	\Psi\left(\frac{L}{2},y\right)=
	\Psi\left(-\frac{L}{2},y\right)\exp(2\pi i\phi),
  &
  \label{BC2}\\
  &
  E_x=-(h/eL) \dot\phi. 
  &
  \label{E}
\end{eqnarray}
In this way we account for a weak magnetic field perpendicular 
to the plane $x,y$. 
If its vector potential is chosen to be parallel to the zigzag edge, 
${\bf A}(y)\| {\bf\hat x}$ and to vanish at $y\to\infty$, 
then at $y=0$ the phase $\phi$ exactly equals to the flux 
through the strip in units of $ch/e$. 
For weak magnetic fields, 
the spatial variation of $\phi$ with the coordinate $y$ 
can be neglected, while its adiabatic variation with time 
implies an electric field along ${\bf\hat x}$ given by Eq.~(\ref{E}).  

\section{Dirac fermion edge states on finite zigzag edges}
\label{finite}

\subsection{Green's function of the system}
\label{Green}

To study the spectral properties of zigzag graphene edges 
it is convenient to use the Green's function approach.
The specifics of its implementation to boundary problems in graphene  
is still scarcely covered in literature (e.g. Refs.~\onlinecite{GT07,Burset08}). 
Below we describe in some detail the main calculation steps 
leading to the final result given by Eqs.~(\ref{G}) -- (\ref{Ga}).

We begin by introducing the retarded Green's function 
$G^R({\bf r}t,{\bf r}^\prime t^\prime)$ 
as a $4\times 4$ matrix in space of the valley (isospin) and the sublattice degrees of freedom 
whose matrix elements are given by
\begin{eqnarray}
&&
G^R_{\alpha\beta}({\bf r}t,{\bf r}^\prime t^\prime)=\frac{\Theta(t-t^\prime)}{i\hbar}\times
\label{G_def}\\
&&
\times
\left<
\psi_{\alpha}({\bf r}t)\psi^\dagger_{\beta}({\bf r}^\prime t^\prime)+
\psi^\dagger_{\beta}({\bf r}^\prime t^\prime)\psi_{\alpha}({\bf r}t)
\right>,
 \nonumber
\end{eqnarray} 
where the brackets $\left<...\right>$ denote averaging with the equilibirum statistical operator 
and the indices $\alpha$ and $\beta$ independently run through all possible combinations 
of the isospin and sublattice indices: $\alpha,\beta=A_+,A_-,B_+,B_-$.
As the zigzag edge [Eq.~(\ref{BC1})] possesses the isospin rotation symmetry generated by $I_3$ (i.e. does not couple the valleys), 
$G^R$ can be decomposed into the direct product: 
\begin{equation}
G^R({\bf r}t,{\bf r}^\prime t^\prime)=\frac{1}{2}\sum_{\tau=\pm 1}
\left(\tau_0 +\tau \tau_3\right)\otimes
G^R_\tau({\bf r}t,{\bf r}^\prime t^\prime)
\label{G_product}
\end{equation}
where $\tau=\pm 1$ labels the valleys (i.e. the two isospin projections) and 
\begin{eqnarray}
	G^R_\tau({\bf r}t,{\bf r}^\prime t^\prime)=
\left(
\begin{array}{cc}
G_{AA|\tau}({\bf r}t,{\bf r}^\prime  t^\prime)  &  G_{AB|\tau}({\bf r}t,{\bf r}^\prime  t^\prime)\\
G_{BA|\tau}({\bf r}t,{\bf r}^\prime  t^\prime)  &  G_{BB|\tau}({\bf r}t,{\bf r}^\prime  t^\prime) 
\end{array}
\right)
\end{eqnarray}
is the matrix Green's function in sublattice space. 
Its time Fourier transform satisfies the equation 
\begin{eqnarray}
(\epsilon\sigma_0 - v\tau\mbox{\boldmath$\sigma$}{\bf p})
G^R_\tau({\bf r},{\bf r}^\prime)=
\sigma_0
\delta( {\bf r}-{\bf r}^\prime ).
\label{DG}
\end{eqnarray}
In terms of $G^R_\tau({\bf r},{\bf r}^\prime)$ the boundary conditions, Eqs.~(\ref{BC1}) and (\ref{BC2}) of the previous section,  read 
\begin{eqnarray}
&
	G^R_\tau=\left. (\mbox{\boldmath$\sigma$}{\bf l}_\tau)G^R_\tau \right|_{y=0},\quad {\bf l}_\tau\equiv {\bf l}_\pm,
&
\label{BCG1}\\
&
        G^R_\tau|_{x=L/2}=G^R_\tau|_{x=-L/2}\exp(2\pi i\phi).
&
\label{BCG2}
\end{eqnarray}
The solution to Eq.~(\ref{DG}) can be sought in the form
\begin{eqnarray}
&&
G^R_\tau({\bf r},{\bf r}^\prime)=
\left( 
\sigma_0+
\frac{v\tau}{\epsilon}\mbox{\boldmath$\sigma$}{\bf p} 
\right)\times
\label{Green1}\\
&&
\times
\sum_{n\in \mathbb{Z}}
\left(
\begin{array}{cc}
G_{AA|\tau k_n}(y,y^\prime) & 0\\
0 & G_{BB|\tau k_n}(y,y^\prime)
\end{array}
\right)
\frac{ {\rm e}^{ik_n(x-x^\prime)} }{L},
\nonumber
\end{eqnarray}
where the diagonal matrix elements 
are the Green's functions on sublattices $A,B$ . 
They are expanded in plane waves ${\rm e}^{ik_n x}$ 
with the wave number
 \begin{equation}
	k_n=(2\pi/L)(n+\phi), \quad n \in  \mathbb{Z}\, (0,\pm 1,...),
	\label{k}
\end{equation}
given by the boundary condition, Eq.~(\ref{BCG2}). 
For $G_{AA,BB|\tau k_n}(y,y^\prime)$ one has the ordinary differential equation, 
\begin{equation}
(\partial^2_y-q^2_n)G_{AA,BB|\tau k_n}(y,y^\prime)
	=\frac{\epsilon}{\hbar^2v^2}\delta(y-y^\prime),
\label{DGA}
\end{equation}
and the boundary conditions following from Eq.~(\ref{BCG1}):
\begin{eqnarray}
  \partial_yG_{AA|\tau k_n}=
  \left.
  \left[
  \frac{ \tau\epsilon (1 - l_{z\tau}) }
       {\hbar v l_{x\tau} } - k_n 
  \right]
  G_{AA|\tau k_n}
  \right|_{y=0},
  \label{BCA}
\end{eqnarray}
\begin{eqnarray}
  \partial_yG_{BB|\tau k_n}=
  \left.
  \left[
  \frac{-\tau \epsilon (1 + l_{z\tau})  }
       { \hbar v l_{x\tau} } + k_n
  \right]
  G_{BB|\tau k_n}
  \right|_{y=0},
  \label{BCB}
\end{eqnarray}
where $q_n=\sqrt{ k_n^2-\epsilon^2/\hbar^2 v^2 }$.
We seek the solution (finite at $y\to\infty$)  in the form 
\begin{equation}
	G_{AA,BB|\tau k_n}(y,y^\prime)=C_{A,B}(y^\prime){\rm e}^{-q_ny}
-\frac{\epsilon}{2\hbar^2 v^2q_n}{\rm e}^{ -q_n|y-y^\prime| },
\nonumber
\end{equation}
where the first term is the solution of the homogeneous equation (\ref{DGA}) 
and the second one is the Green's function of the unbounded system.   
The coefficients $C_{A,B}$ are obtained from Eqs.~(\ref{BCA}) 
and (\ref{BCB}) with the following results:
\begin{eqnarray}
&&
G_{AA|\tau k_n}(y,y^\prime)=\frac{\epsilon}{2\hbar^2v^2q_n}
\left(
{\rm e}^{-q_n(y+y^\prime)} - {\rm e}^{ -q_n|y-y^\prime| }
\right)
\nonumber\\
&&
+
\frac{ (1+l_{z\tau})(q_n + k_n) - \tau \epsilon l_{x\tau}/\hbar v }
{ 2( \epsilon - \hbar v\tau k_n l_{x\tau}+i0) }
\,{\rm e}^{-q_n(y+y^\prime)},
\label{GA}
\end{eqnarray}
\begin{eqnarray}
&&
G_{BB|\tau k_n}(y,y^\prime)=\frac{\epsilon}{2\hbar^2v^2q_n}
\left(
{\rm e}^{-q_n(y+y^\prime)} - {\rm e}^{ -q_n|y-y^\prime| }
\right)
\nonumber\\
&&
+
\frac{ (1-l_{z\tau})(q_n - k_n) + \tau \epsilon l_{x\tau}/\hbar v }
{ 2( \epsilon - \hbar v\tau k_n l_{x\tau} +i0) }
\,{\rm e}^{-q_n(y+y^\prime)},
\label{GB}	
\end{eqnarray}

The results of the above calculations can be summarized in the expression for  
the full matrix Green's function:
\begin{eqnarray}
&&
G^R({\bf r},{\bf r}^\prime)=\sum_{\tau=\pm 1, n\in \mathbb{Z}}
\left( \frac{\tau_0 + \tau\tau_3}{2} \right)\otimes
\left( \sigma_0 +\frac{v\tau}{\epsilon}\mbox{\boldmath$\sigma$}{\bf p} \right)
\nonumber\\
&&
\times
\left( 
G^{s}_{\tau k_n}(y,y^\prime)\sigma_0 + 
G^{a}_{\tau k_n}(y,y^\prime)\sigma_z 
\right)
\frac{{\rm e}^{ik_n(x-x^\prime)} }{L}.
\label{G}
\end{eqnarray}
We introduce the symmetric 
$G^{s}_{\tau k_n}(y,y^\prime)=(G_{AA|\tau k_n} + G_{BB|\tau k_n})/2$ 
and asymmetric $G^{a}_{\tau k_n}(y,y^\prime)=(G_{AA|\tau k_n} - G_{BB|\tau k_n})/2$ 
sublattice functions given explicitly by
\begin{eqnarray}
&&
G^{s}_{\tau k_n}(y,y^\prime)=
\frac{\epsilon}{2\hbar^2v^2q_n}
\left(
{\rm e}^{-q_n(y+y^\prime)} - {\rm e}^{ -q_n|y-y^\prime| }
\right)
\nonumber\\
&&
+
\frac{ q_n + k_nl_{z\tau} }{ 2( \epsilon - \hbar v\tau k_nl_{x\tau} +i0) }
\,{\rm e}^{-q_n(y+y^\prime)}, 
\label{Gs}
\end{eqnarray}
\begin{eqnarray}
G^{a}_{\tau k_n}(y,y^\prime)=
\frac{k_n + q_nl_{z\tau} -\tau\epsilon l_{x\tau}/\hbar v  }
{2( \epsilon - \hbar v\tau k_n l_{x\tau} +i0)}
\,{\rm e}^{-q_n(y+y^\prime)}.
\label{Ga}
\end{eqnarray} 
In the equations above the denominators vanish at 
$\epsilon=\hbar v\tau k_nl_{x\tau}$. 
To identify this as a pole, we should make sure that the nominators 
remain finite as $ \epsilon\to\hbar v\tau k_nl_{x\tau}$.
In this limit, the Green's function (\ref{G}) behaves as
\begin{eqnarray}
&&
G^R({\bf r},{\bf r}^\prime)\approx 
-\frac{1}{2L}\sum_{\tau=\pm 1, n\in \mathbb{Z}}
\left( \tau_0 + \tau\tau_3 \right)\otimes 
(\sigma_0 + \mbox{\boldmath$\sigma$}{\bf l}_\tau)
\nonumber\\
&&
\times
\frac{\Theta( k_nl_{z\tau} )}{\epsilon  - \hbar v\tau k_nl_{x\tau} +i0 } 
\partial_y\,{\rm e}^{-|k_nl_{z\tau}|(y+y^\prime) + ik_n(x-x^\prime)},
\label{G1}
\end{eqnarray} 
and we can see that the pole exists only 
if the unit step function $\Theta( k_nl_{z\tau} )$ is not zero:
\begin{equation}
\epsilon_{\tau, n}=\hbar v\tau k_nl_{x\tau},
\qquad 
k_n l_{z\tau}>0.
\label{Spectrum}
\end{equation}    
This is the spectrum of the states, 
decaying exponentially from the edge $y=0$ 
and propagating along $x$.

\subsection{Edge-state spectrum, isospin-orbit coupling and zero modes}
\label{edge}

\begin{figure}[t]
\begin{center}
\epsfxsize=0.9\hsize
\epsffile{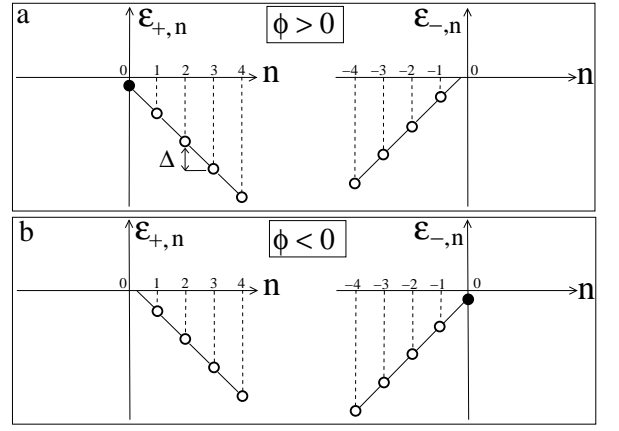}
\end{center}
\caption{
Edge states in $+$ and $-$ valleys, Eq.~(\ref{Spectrum1}) 
for (a) positive and (b) negative magnetic flux $\phi$.
We assume the zigzag confinement parameters, $l_x<0$ and $l_z>0$, 
so that the edge states exist below the Fermi level as inferred 
from tunneling spectroscopic measurements~\cite{Koba05,Niimi06}.
The external magnetic flux shifts 
the levels in $+$ and $-$ valleys in the opposite directions 
such that the zero mode $n=0$ (filled circle) occurs only 
in one of the valleys: $+$ one for $\phi>0$ and $-$ one for $\phi<0$.  
The valley-dependent zero mode violates Kramers' symmetry of 
the edge-state spectrum for arbitrary small $\phi$. 
}
\label{spectr}
\end{figure}

Let us analyze the edge-state spectrum, Eq.~(\ref{Spectrum}) 
in some more detail.
With the requirements of the $x$-parity and time-reversal symmetry 
[see, Eq.~(\ref{n})] and for $k_n$ given by Eq.~(\ref{k}), 
we have 
\begin{eqnarray}
&
	\epsilon_{\tau, n}={\rm sgn}(l_x)\Delta\, \tau (n+\phi) ,
\qquad 
 \tau (n+\phi)l_{z}>0,
&
\label{Spectrum1}\\
&
\Delta=hv|l_x|/L,\qquad n=0,\pm 1...
&
\label{Delta}
\end{eqnarray}
It is equidistant with the level spacing $\Delta$ 
and particle-hole asymmetric because of the restriction 
$(n+\phi)\tau l_{z}>0$ [see, Fig.~\ref{spectr}].
The phase $\phi$ results in the homogeneous shift of the levels.
Let us consider $|\phi|\ll 1$ and neglect the shift in all of the states 
except the zero mode $n=0$: 
\begin{eqnarray}
&&
	\epsilon_{\tau, n}={\rm sgn}(l_x)\Delta\, \tau n,
\quad 
 \tau n \, l_z>0,\quad n=\pm 1,... 
\label{Spectrum2}\\
&&
\epsilon_{\tau, 0}={\rm sgn}(l_x)\Delta\, \tau\phi ,
\quad 
 \tau \phi \, l_z>0,\quad n=0.
\label{Spectrum3}
\end{eqnarray}
We see that the states with $n=\pm 1,...$ exhibit Kramers' symmetry 
under $\tau, n\to - \tau, -n$  
resulting from the coupling between the valley (isospin) 
degree of freedom $\tau$ and the orbital quantum number $n$. 
{\em The isospin-orbit coupling originates 
from the broken isospin rotation symmetry 
discussed in Sec.~\ref{zig}. 
The coupling constants are given by the parameters 
$\,l_x$ and $l_z$ of the zigzag confinement}.
For the hard-wall zigzag edge ($l_x=0$), 
we find the degenerate zero-energy state $\epsilon_{\tau, n}=0$. 
This is in agreement with the tight-binding calculations 
for zigzag graphene ribbons
(e.g. Refs.~\onlinecite{Fujita96,Waka00}) 
if their results are extrapolated to the case of the infinite width
when the edges become isolated.   

The zero mode, Eq.~(\ref{Spectrum3}) stands out 
because it is due to the coupling between the isospin 
and the electromagnetically induced momentum  
$k_0=(2\pi/L)\phi$. {\em This mode breaks the Kramers' symmetry 
of the edge-state spectrum since it exists only for one of 
the isospin projections $\tau={\rm sgn}(\phi l_z)$, 
i.e. only in one of the valleys}. 
In other words, there is a valley polarization effect. 
It is studied quantitatively in the next section.

\section{Valley polarization}
\label{valley}
\subsection{Spontaneous polarization}
\label{spont}

To quantitatively characterize the valley polarization effect 
we introduce the local isospin polarization:
\begin{eqnarray}
	p(\epsilon,{\bf r})
	&=&
	-\frac{1}{\pi}\,{\rm Im\, Tr}\, I_3\, G^R({\bf r},{\bf r})
	=
	\nonumber\\
  &=&
  -\frac{2}{\pi L}\sum\limits_{\tau=\pm 1, n\in\mathbb{Z}}
  \frac{\tau}{2}\,{\rm Im}\,G^s_{\tau k_n}(y,y),
\label{LIP}
\end{eqnarray}
where ${\rm Im}$ denotes the imaginary part, 
the trace ${\rm Tr}$ of the Green's function (\ref{G}) is taken in 
$\tau\otimes\sigma$ space, 
and the function $G^s_{\tau k_n}(y,y)$ is given by Eq.~(\ref{Gs}) 
of the previous section. As we are interested in 
the edge isospin polarization, we relevant contribution to 
${\rm Im}\,G^s_{\tau k_n}(y,y)$ comes from the pole in Eq.~(\ref{Gs}): 
\begin{eqnarray}
 	p_e(\epsilon,y)=
 	-\frac{1}{L}\,\partial_y\sum\limits_{\tau, n\in\mathbb{Z}}
 	&\frac{\tau}{2}&
 	{\rm e}^{ -2|k_n l_z|y }\Theta( k_n \tau l_z )\times
\nonumber\\
&\times&
  \delta( \epsilon - \epsilon_{\tau, n} ).\,\,\,
\label{LIP_e}
\end{eqnarray}
Note that for the zero mode the step function 
$\Theta( \phi \tau l_z )$ indicates the breaking of the Kramers' symmetry. 

Next we calculate the zero-temperature ground-state isospin density 
localized at the edge as
\begin{eqnarray}
	&&
	i_e(y)=\int^0_{-\infty}d\epsilon\, p_e(\epsilon,y)=
	\label{i_def}\\
	&&
	=-\frac{1}{L}\partial_y\sum\limits_{\tau, n\in\mathbb{Z}}
	\frac{\tau}{2}\,
 	{\rm e}^{ -2|k_n l_z|y }
 	\Theta(k_n \tau l_z)\Theta(-k_n \tau l_x)
  \label{i1}\\
  &&
  =-\frac{\Theta(-l_xl_z)}{L}\partial_y\sum\limits_{\tau, n\in\mathbb{Z}}
	\frac{\tau}{2}\,
 	{\rm e}^{ -2|k_n l_z|y }
 	\Theta(k_n \tau l_z).
 	\label{i2}
\end{eqnarray}
To obtain the last formula we used the identity 
$\Theta(x)\Theta(y)=\Theta(x)\Theta(xy)$.
The summations over $\tau=\pm 1$ and $n$ can be done exactly: 
\begin{eqnarray}
	&&
	i_e(y)
	=-\frac{N}{2L}
	\partial_y
	\sum\limits_{n=-\infty}^{\infty}
	{\rm sgn}(n+\phi)
 	{\rm e}^{ -|n + \phi|y/\lambda }
  \label{i3}\\
  &&
  =-\frac{N}{2L}
	\partial_y
	\left[
	{\rm sgn}\,\phi\,{\rm e}^{-|\phi|y/\lambda}
  -
  \frac{2\sinh(\phi\, y/\lambda)}{{\rm e}^{\,y/\lambda} -1}
  \right],
  \label{i4}\\
  &&
  N=\Theta(-l_xl_z){\rm sgn}\,l_z, \qquad \lambda=\frac{L}{4\pi |l_z|}.
  \label{N}
\end{eqnarray}
In Eq.~(\ref{i4}) the first term, nonanalytic in $\phi$, 
is due to the zero edge mode $n=0$. 
Its penetration length depends on the flux $\phi$ and diverges at $\phi\to 0$.
The second term accounts for the rest of the edge states $n=\pm 1,...$ 
It is an analytic function of $\phi$. The edge-state penetration length 
is measured in units of $\lambda$ given in Eq.~(\ref{N}), 
and the flux is confined to a one-period interval chosen as $-1/2<\phi< 1/2$.

We note that depending on the boundary parameters 
the factor $N$ (\ref{N}) takes integer values $0$ or $\pm 1$.  
The case $N=0$ corresponds to edge states 
above the Fermi level $\epsilon=0$, for which the zero-temperature 
occupation number $\Theta(-l_xl_z)=0$.  
In what follows, we focus on the opposite situation, 
i.e. the edge states below the Fermi level and 
\begin{eqnarray}
	N={\rm sgn}\,l_z=\pm 1, \quad l_xl_z<0,
\label{N1}
\end{eqnarray}
which is supported by the tunneling spectroscopy~\cite{Koba05,Niimi06}.

Finally, we obtain the total isospin carried by the edge states as
\begin{eqnarray}
	I_e=L\int^{\infty}_{0}dy\, i_e(y)
     =\left( \frac{{\rm sgn}\,\phi}{2} -\phi \right){\rm sgn}\,l_z.
\label{Ie}
\end{eqnarray}
The zero mode results in the discontinuity at $\phi=0$, 
due to which  in the limit $\phi\to 0$ the total isospin remains finite (half-integer): 
\begin{eqnarray}
I_e=\frac{1}{2}\,{\rm sgn}\,(\phi l_z), \quad  \phi\to 0.
\label{Ie1}
\end{eqnarray}
This implies that {\em the ground state does not share 
the time-reversal symmetry of the original equations~(\ref{DG}) -- (\ref{BCG2}) in the limit  $\phi\to 0$.  
In this sense, the zero mode violates the time-reversal and Kramers' symmetries spontaneously, 
with the resulting spontaneous valley polarization.}

\subsection{Valley Hall polarization}
\label{Hall_pol}

The accumulated isospin, Eq.~(\ref{Ie}) contains    
a linear term $\propto\phi$. It comes from the Kramers' degenerate 
edge states with $n=\pm 1,...$ [see, Eq.~(\ref{Spectrum2})]. 
Such a property of Kramers' degenerate edge states was first 
noticed in the theory of quantum spin Hall systems 
(e.g. Refs.~\onlinecite{Kane05,Sheng05,Bernevig06,Koenig07}). 
The recent interest in these systems is motivated by 
the principal possibility to realize a time-reversal invariant 
integer quantum Hall state in which the spin Hall conductance is quantized. 
From Eq.~(\ref{Ie}) it is possible to derive the analogue of 
the quantum spin Hall conductance. 
Let us calculate the isospin current as   
the rate of adiabatic change of the isospin: 
$\dot I_e =\dot\phi\, \partial I_e/\partial\phi=
-(eE_xL/h)\partial I_e/\partial\phi$, 
which assumes $\phi\not= 0$ and Eq.~(\ref{E}). 
The derivative $\dot I_e$ gives the transverse isospin flow 
in response to the voltage drop $E_xL$ along the edge: 
\begin{equation}
	\dot I_e=G_{iH}E_xL,\qquad G_{iH}=\frac{e}{h}{\rm sgn}\,l_z,
	\label{iH}
\end{equation}
where the quantum isospin Hall conductance $G_{iH}$ 
takes the universal values $\pm e/h$. 
As the zigzag-terminated graphene supports 
the edge states without any excitation gap in the bulk, 
the conductance (\ref{iH}) is hardly the signature of 
any bulk topological order~\cite{Discussion}. 
We rather interpret it as the measure of the valley polarization rate 
at the edges.

\section{Signatures of the valley polarization in observables}
\label{signs}

\subsection{Tunneling density of states}
\label{tunnel}

\begin{figure}[t]
\begin{center}
\epsfxsize=0.8\hsize
\epsffile{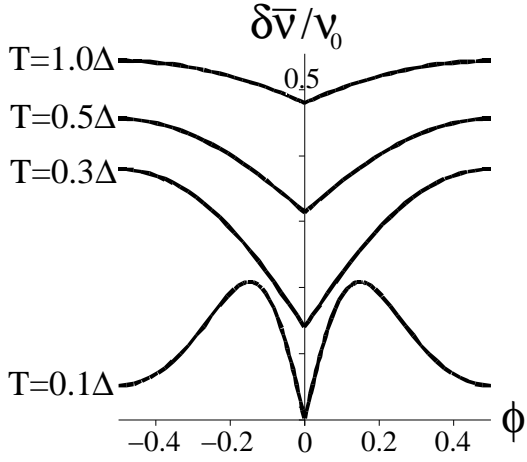}
\end{center}
\caption{
Tunneling density of states, Eq.~(\ref{TDOS1}) vs. magnetic flux (in units of $ch/e$)
at various temperatures; $y=0.5\lambda$, $\nu_0=1/\Delta L\lambda$. 
}
\label{nu_phi}
\end{figure}

The presence of the valley polarization can be inferred 
from the magnetic field dependence of the zero mode. 
One of the possibilities is to measure the local tunneling 
conductance in the presence of a weak magnetic.   
At zero bias and the finite temperature $T$, 
the tunneling conductance is proportional to the tunneling 
density of states
\begin{equation}
	\overline\nu(T,\phi)=\int_{-\infty}^{\infty}
	d\epsilon\, \left(-\frac{\partial f(\epsilon, T)}{\partial\epsilon}\right)
	\nu(\epsilon,\phi),
	\label{TDOS}
\end{equation}
which is the convolution of the local spectral density of states, 
$\nu(\epsilon,\phi)$ and the energy derivative of 
the Fermi distribution function, $f(\epsilon, T)$.
The local spectral density of states is obtained 
from the Green' function, Eq.~(\ref{G}) as
\begin{eqnarray}
	\nu(\epsilon,{\bf r})
	&=&-\frac{1}{\pi}\,{\rm Im\, Tr}\,G^R({\bf r},{\bf r})=
	\nonumber\\
	&=&-\frac{2}{\pi L}\sum\limits_{\tau=\pm 1, n\in\mathbb{Z}}
	{\rm Im}\,G^s_{\tau k_n}(y,y).
\label{LDOS}
\end{eqnarray}
The edge-state contribution to ${\rm Im}\,G^s_{\tau k_n}(y,y)$ 
comes from the pole in Eq.~(\ref{Gs}):
\begin{eqnarray}
 	\nu_e(\epsilon,y)=
 	-\frac{1}{L}\partial_y\sum\limits_{\tau, n\in\mathbb{Z}}
 	{\rm e}^{ -2|k_n l_z|y }
 	\Theta(k_n \tau l_z)
  \delta( \epsilon - \epsilon_{\tau, n} ).\,\,\,
\label{nu_e}
\end{eqnarray}
From Eqs.~(\ref{TDOS}) and (\ref{nu_e}) 
one can obtain the edge-state contribution 
to the tunneling density of states as
\begin{eqnarray}
&&
		\delta\overline\nu(T,\phi)=\frac{1}{4LT\lambda}
		\biggl\{
     \frac{|\phi|{\rm e}^{-|\phi|y/\lambda}}
     {\cosh^2\left( \frac{\phi\Delta}{2T}\right)}+
\label{TDOS1}\\
&&
     +\sum_{n=1}^{\infty}
     \biggl[
     \frac{(n+\phi){\rm e}^{-\phi\,y/\lambda}}
     {\cosh^2\left( \frac{(n+\phi)\Delta}{2T}\right)}
     +
     \frac{(n-\phi){\rm e}^{\phi\,y/\lambda}}
     {\cosh^2\left( \frac{(n-\phi)\Delta}{2T}\right)}
     \biggr]{\rm e}^{-n\,y/\lambda}
     \biggr\},
     \nonumber
\end{eqnarray}
where $\Delta$ is the level spacing given by Eq.~(\ref{Delta}). 

\begin{figure}[t]
\begin{center}
\epsfxsize=0.7\hsize
\epsffile{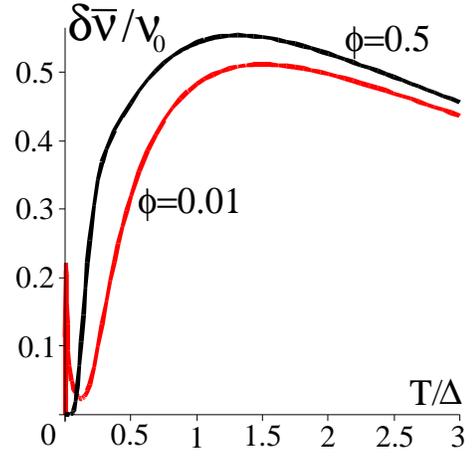}
\end{center}
\caption{(Color online)
Tunneling density of states, Eq.~(\ref{TDOS1}) vs. temperature (in units of level spacing $\Delta$) 
for small ($\phi=0.01$) and large ($\phi=0.5$) flux values; $y=0.5\lambda$, $\nu_0=1/\Delta L\lambda$. 
}
\label{nu_T}
\end{figure}

Figure~\ref{nu_phi} shows that the flux dependence of $\delta\overline\nu$ 
is nonanalytic, indicating the spontaneous valley polarization at 
$|\phi|\to 0$. 
The nonanalyticity is present in a wide range of temperatures. 
The reason is that the zero-mode term always dominates 
the flux dependence near $\phi=0$ because it is linear in $|\phi|$, 
while the rest of the sum varies as  $\phi^2$.
As demonstrated in Fig.~\ref{nu_T}, 
for small $\phi=0.01$ (red curve) the zero-mode also dominates 
the low-temperature behavior of  $\delta\overline\nu$, 
showing a $1/T$ increase when  $T$ becomes much smaller 
than the level spacing $\Delta$.    
This feature is due to the fact that for $\phi\ll 1$ 
the energy of the zero mode $\propto \phi\Delta \ll \Delta$.  
In contrast, for the rest of the sum in Eq.~(\ref{TDOS1})
the relevant energy scale is set by the level spacing $\Delta$. 

\subsection{Local electric Hall conductivity}
\label{local_Hall}

Although the zero-mode behavior in the tunneling density of states 
signals the valley polarization effect, 
this observable does not provide the direct access 
to the accumulated isospin. Here we intend to show 
that the accumulated isospin is directly related to 
the local electric Hall conductivity.  

\begin{figure}[t]
\begin{center}
\epsfxsize=1\hsize
\epsffile{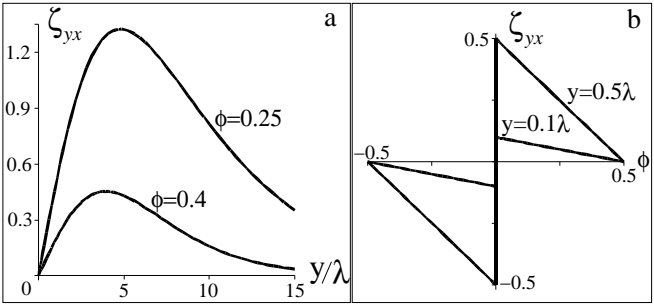}
\end{center}
\caption{
Hall conductivity in units of $e^2/h$, Eq.~(\ref{Hall}) vs. 
(a) position and (b) dimensionless magnetic flux.
}
\label{sigma}
\end{figure}

As the first step, we use Eq.~(\ref{nu_e}) to 
calculate the zero-temperature ground-state charge density 
localized at the edge:
\begin{eqnarray}
	&&
	\rho_e(y)=e\int^0_{-\infty}d\epsilon\, \nu_e(\epsilon,y)=
	\label{rho_def}\\
	&&
	=-\frac{e}{L}\partial_y\sum\limits_{\tau, n\in\mathbb{Z}}
 	{\rm e}^{ -2|k_n l_z|y }
 	\Theta(k_n \tau l_z)\Theta(-k_n \tau l_x)
  \label{rho1}\\
  &&
  =-\frac{e\Theta(-l_xl_z)}{L}\partial_y\sum\limits_{\tau, n\in\mathbb{Z}}
 	{\rm e}^{ -2|k_n l_z|y }
 	\Theta(k_n \tau l_z).
 	\label{rho2}
\end{eqnarray}
Again the summations over $\tau=\pm 1$ and $n$ 
can be done explicitly:
\begin{eqnarray}
&&
	\rho_e(y)=-\frac{e\Theta(-l_xl_z)}{L}\partial_y
  \sum\limits_{n=-\infty}^{\infty}
  {\rm e}^{ -|n + \phi|y/\lambda}=
  \label{rho3}\\
&&
  =-\frac{e\Theta(-l_xl_z)}{L}\partial_y
  \left[
  {\rm e}^{-|\phi|y/\lambda}
  +
  \frac{2\cosh(\phi\, y/\lambda)}{{\rm e}^{\,y/\lambda} -1}
  \right],
  \label{rho4}
\end{eqnarray}
with $|\phi|\leq 1/2$. The $\phi$ dependence of Eq.~(\ref{rho4}) 
allows us to take the adiabatic time derivative,
$\dot\rho_e=\dot\phi\, \partial_\phi\rho_e 
=-(eE_xL/h)\, \partial_\phi\rho_e$, 
and obtain the following continuity equation:
\begin{equation}
\dot\rho_e=-\partial_yj_y, \qquad j_y=\varsigma_{yx}E_x,
\label{Con}
\end{equation}
where $j_y$ is the Hall current density induced by 
the transverse electric field $E_x$, and 
$\varsigma_{yx}$ is the local {\em position-dependent} 
Hall conductivity given by 
\begin{equation}
\varsigma_{yx}(y,\phi)=\frac{4e^2l_z}{\hbar} y\int^\infty_ydy^\prime\,i_e(y^\prime,\phi).
\label{Hall}
\end{equation}
It is expressed in terms of the edge isospin density given by Eq.~(\ref{i4}).
The existence of the electric current density, $j_y$ normal 
to the system's boundary is consistent 
with the charge conservation because 
at the edge $y=0$ the conductivity $\varsigma_{yx}(0,\phi)$ is zero 
[see, also, Fig.~\ref{sigma}(a)]. 
It also vanishes far away from the edge: 
$\varsigma_{yx}(y\to\infty,\phi)\to 0$, so that 
the total edge charge is conserved: $\int^\infty_0dy\,\dot\rho_e=0$.    

At distances smaller than the characteristic penetration length, 
$y\ll\lambda$, the Hall conductivity is simply proportional 
to the total isospin carried by the edge states:
\begin{eqnarray}
	\varsigma_{yx}(y,\phi)\approx
	\frac{2e^2{\rm sgn}\,l_z}{h}\frac{y}{\lambda}\,I_e(\phi)=
	\frac{2e^2}{h}\frac{y}{\lambda}
	\left( \frac{{\rm sgn}\,\phi}{2} -\phi \right),\,
	\label{Hall1}
\end{eqnarray}
showing the same nonanalytic flux dependence as $I_e(\phi)$ in Eq.~(\ref{Ie}) 
[see, Fig.~\ref{sigma}(b)].

\begin{figure}[t!]
\begin{center}
\epsfxsize=0.8\hsize
\epsffile{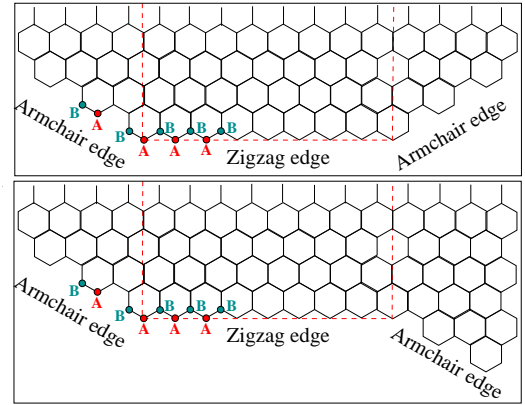}
\end{center}
\caption{(Color online)
Other realizations of a finite-length zigzag boundary between two armchair edges. 
}
\label{zigzags}
\end{figure}

In conclusion, we discuss the applicability of the results of this paper. 
First of all, the lattice prototype of our continuous model [Fig.~\ref{Geo}(a)]
is only one of many possible realizations of a finite-length zigzag boundary. 
It is nevertheless clear that for the honeycomb lattice 
an armchair/zigzag/armchair edge structure
is rather typical [see, e.g. Fig.~\ref{zigzags}].  
Independently of its concrete realization, 
the edge states  must experience multiple intervalley backscattering 
from the two armchair regions, resulting in the bound states. 
The experimental estimate~\cite{Niimi06} of the typical length 
of zigzag edges is of order of $10\,{\rm nm}$. 
This is large enough for the applicability of our continuum model 
and, on the other hand, is shorter than the typical mean free path in graphene, 
which is required for the ballistic quantization.  
For samples with longer edges, multiple electron scattering due to 
boundary and bulk disorder may come into play, 
as revealed by recent numerical studies~\cite{Evaldsson08,Mucciolo08}.   
The quantization effects studied in this paper 
are characteristic to isolated zigzag edges 
as opposed to the size-quantization in zigzag graphene ribbons. 
Because the control over graphene edges is 
still a serious experimental issue,  
it seems easier to obtain graphene samples 
with isolated zigzag edges rather than to produce 
zigzag-terminated ribbons. 
In this sense, local tunneling spectroscopy 
is currently the most adequate tool for investigating the edge states 
in graphene. As for the results on 
the local electric Hall conductivity (\ref{Hall}), 
their verification may present a challenging experimental task. 
It should however be achievable with increasing control over 
the boundary effects in graphene.

\acknowledgements
The author thanks F. Guinea, M. Hentschel and M. I. Katsnelson for discussions. 
The work was supported the Emmy-Noether Programme of 
the German Research Foundation (DFG).



\begin{thebibliography}{00}
\bibitem{Discussion}
This issue requires a separate discussion that will be presented elsewhere.
\end{thebibliography}
\end{document}